\documentclass{article}

\usepackage{arxiv}

\usepackage[utf8]{inputenc} % allow utf-8 input
\usepackage[T1]{fontenc}    % use 8-bit T1 fonts
\usepackage{hyperref}       % hyperlinks
\usepackage{url}            % simple URL typesetting
\usepackage{booktabs}       % professional-quality tables
\usepackage{amsfonts}       % blackboard math symbols
\usepackage{nicefrac}       % compact symbols for 1/2, etc.
\usepackage{microtype}      % microtypography
\usepackage{lipsum}
\usepackage{graphicx}

\title{Electrically reconfigurable silicon photonic filter enabled by embedded phase change material in microring resonator}

\author{
  Nadir Ali \\
  Department of Physics\\
  Indian Institute of Technology Roorkee\\
  Roorkee, India - 247667 \\
  \texttt{nali@ph.iitr.ac.in} \\
   \And
 Rajesh Kumar \\
  Department of Physics\\
  Indian Institute of Technology Roorkee\\
  Roorkee, India - 247667 \\
  \texttt{rajeshfph@iitr.ac.in} \\
}

\begin{document}
\maketitle

\begin{abstract}
We report a tunable optical filter based on phase change material $Ge_{2}Sb_{2}Te_{5}$ embedded in a silicon microring resonator. The high thermo-optic coefficient of Ge\textsubscript{2}Sb\textsubscript{2}Te\textsubscript{5} in amorphous phase enables tuning of resonance wavelength in broad range with a very small active volume. The low-loss indium-tin-oxide electrodes are employed to induce Joule heating in Ge\textsubscript{2}Sb\textsubscript{2}Te\textsubscript{5}-Si active waveguide region. The electrically induced heating in the active region alters the effective refractive index of hybrid microring resulting in a wavelength tuning of 1.04 nm for an applied voltage of only 3V. The device exhibits high extinction ratios in the range of 20-41 dB and a compact active footprint 0.96 $\mu m^{2}$ only and is suitable for the large scale reconfigurable integrated photonic circuits.
\end{abstract}

% keywords can be removed
\section{Introduction}
Silicon photonics has emerged as a preferred platform for the integrated photonics due to its low loss light guidance, high component density and compatibility with CMOS manufacturing technology \cite{b1,b2}. The large-scale photonic integrated circuits (PICs) have been realized recently using silicon photonics platform. Many PICs are designed to be reconfigurable instead of task specific, and can take advantage of mass manufacturing of a single platform \cite{b3}. Most of the Silicon (Si) based reconfigurable devices exploit the refractive index tuning of Si through volatile electro-optic or thermo-optic effect \cite{b4}. However, in such devices a large interaction length is required due to the moderate tunability of Si refractive index. Therefore, these devices suffer from large footprint and high energy consumption, and are not well suited for dense integration from reconfigurability point of view.\\
For on-chip optical communication, miniaturized optical components that can be reconfigurable are essential to realize various optical functionalities with high density of integration. One of the key building-blocks in wavelength division multiplexing (WDM) based high capacity optical communication systems is an optical filter. An optical filter is required to separate a specific wavelength channel from multiple wavelength channels and can be tunable across a range of wavelengths. For silicon platform based on-chip and chip to chip optical communication, optical filters need to have a small footprint, high energy efficiency and compatibility with CMOS platform. \\
%Si based integrated optical filters are mostly based on thermo-optic effect of Si and suffers from large footprint and energy consumption. Optical filter is one of the essential component required for the wavelength division multiplexing (WDM) in an optical communication system.
%The WDM systems provide high bandwidth by sending multiple channels of different wavelength through an optical fiber.
% by application of short optical or electrical pulses
Recently, optically active phase change materials (PCMs) have been utilized to realize different non-volatile integrated photonic devices including waveguide switches \cite{b5}, modulators \cite{b6}, couplers \cite{b7}, and photonic memories \cite{b8}. The reversible switching in these devices is achieved by changing the phase of PCM between its amorphous and crystalline phases, resulting in large change in the electrical and optical constants. Such a non-volatile platform is useful for realizing low energy consuming photonic devices. Among many PCMs, Ge\textsubscript{2}Sb\textsubscript{2}Te\textsubscript{5} (GST) has attracted great attention due to its plethora of properties including high index contrast, sub-nanosecond switching time \cite{b9}, high scalability, CMOS compatibility, and non-volatile nature \cite{b10}. Especially, the drastic change in real and imaginary part of refractive indices of GST between its two phases leads to a large change in optical phase or amplitude within a small active volume. The phase transition can be triggered by heating GST electrically or optically \cite{b11,b12}. In electrically triggered phase transition, current flows through the GST to induce Joule heating \cite{b13}.\\
%Moreover, between the complete amorphous and crystalline phase states transmission, distinct transmission states can be generated by applying control excitation signal
Here, we report on the proof-of-concept investigations through design and simulations of an integrated tunable optical filter using GST. The said tunable filter is based on combining GST in amorphous phase with silicon-on-insulator (SOI) microring resonator. This approach has advantage of having ultra low active volume while still having high extinction ratio and large span wavelength tuning. Our investigations revealed that the proposed filter can have extinction ratios in the range of 20--41 dB and the tuning of wavelength by more than 1000 pm.
\begin{figure*}
	\centering
		\includegraphics[scale=.5]{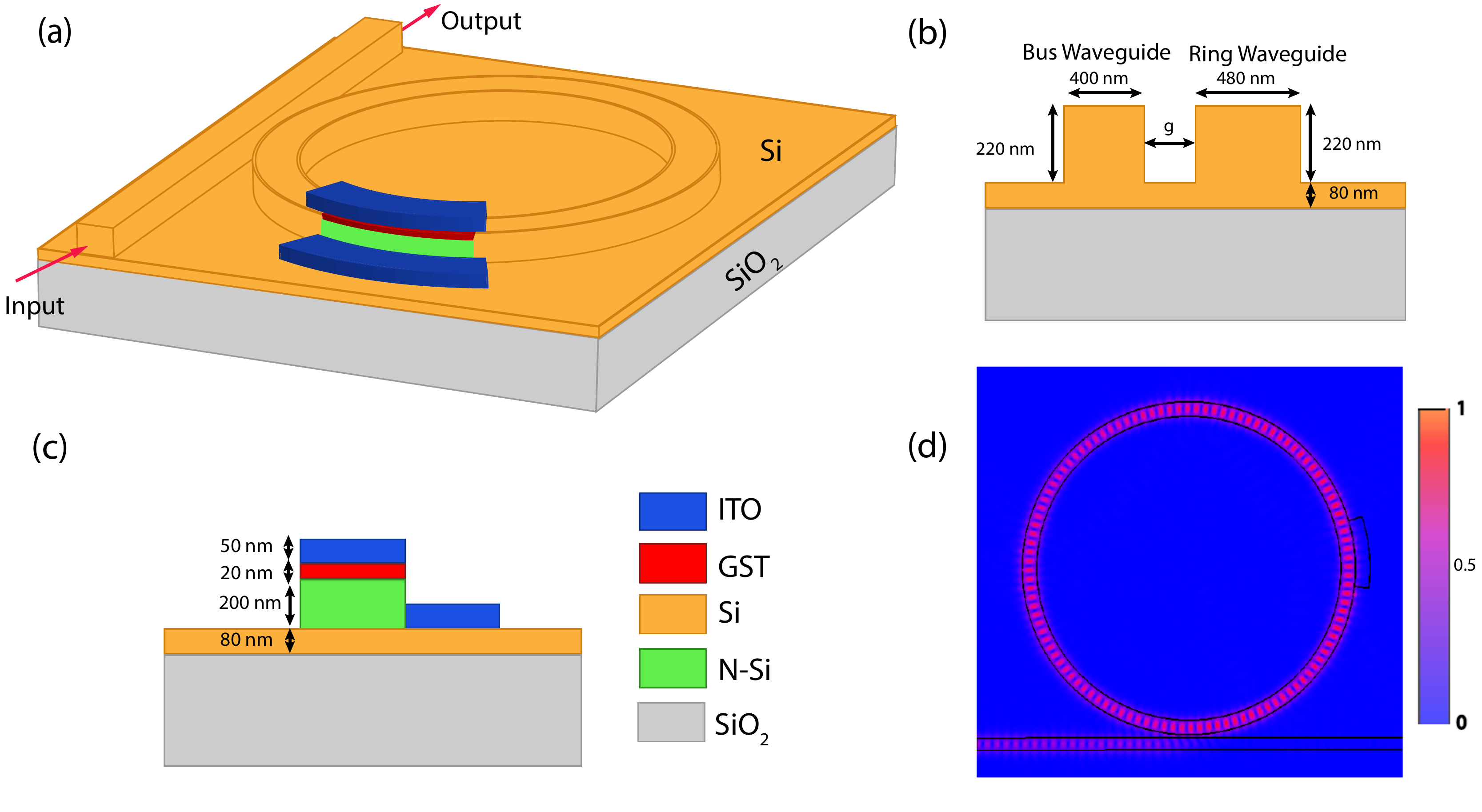}
	\caption{(a) Schematic view of tunable optical filter enabled by embedded amorphous GST in microring resonator, (b) cross-section of the ring and bus waveguides, (c) cross-section of the active region of switch; amorphous GST substituted in partially etched silicon microring and ITO top and botton electrodes in place, (d) electric field profile in the proposed microring resonator.}
	\label{fig1}
\end{figure*}
\section{Concept of Proposed Optical Filter}
%Microring resonators (MRRs) are versatile components and can be used as building-blocks in various integrated optical applications. In optical communication, MRRs show potential to be utilized as a tunable optical filter for wavelength division multiplexing (WDM) system. 
A tunable optical filter can be realized using all pass configuration of microring resonator, which consists of a bus and microring waveguide, and has one input and output (through) port. The light launched through the input port couples to the microring when the resonance condition, $2 \pi R n_{eff} = m \lambda $, is satisfied. In this relation R and $n_{eff}$ denote radius and effective index of microring resonator respectively while m and $\lambda$ are integer and wavelength of light respectively. The transmission spectrum of microring resonator is given by the equation:
\begin{eqnarray*}
    T = \frac{a^{2}-2at \cos{\phi}+t^{2}}{1-2tacos(\phi)+(a t)^{2}} \\
     a = \exp(- \alpha l) 
    %  \text{ and }
    %  \phi = \beta l 
\end{eqnarray*}
 where $l$, $a$, $\alpha$, $t$, and $\phi$ denote circumference of the microring, round trip loss, loss coefficient, straight coupling coefficient, and round trip phase shift inside the microring respectively.
 At resonance $\phi$ = 2m$\pi$, accordingly the transmission spectrum of the microring becomes
 \begin{eqnarray*}
    T = \frac{(a-t)^{2}}{(1-at)^{2}}
\end{eqnarray*}
The thermo-optic effect has been widely used in microring based filters. In this approach, the effective refractive index of Si microring is changed by integrating heater on the platform \cite{b14}, resulting in a shift of resonance wavelength corresponding to the rise in temperature given by:
 \begin{eqnarray*}
     \frac{\Delta \lambda_{res}}{\Delta T} = \frac{\lambda}{n_{g}}\frac{dn}{dT}
 \end{eqnarray*}
 Where $\lambda_{res}$, $n_{g}$, and $dn/dT$ denote resonance wavelength, group index, and thermo-optic coefficient respectively. The shift in the resonance wavelength is directly proportional to the thermo-optic coefficient of the material.\\
 The Si thermo-optic coefficient is $ 1.8 \times 10^{-4} K^{-1}$ \cite{b15}. In the device reported here, we exploit the thermo-optic coefficient of amorphous GST ($\sim 1.1 \times 10^{-3} K^{-1}$) \cite{b16} which is one order of magnitude higher than the Si. Therefore, high wavelength response can be achieved by combining GST with Si microring. Since GST has a large extinction coefficient placing it in a complete Si microring will lead to unacceptably large propagation loss. To avoid large propagation loss, and simultaneously achieve large tuning of resonance wavelength, we embedded GST in a small portion of the Si microring as shown in Fig.~\ref{fig1}(a). The embedding of GST enhances its interaction with guided mode, and allows proper utilization of its thermo-optic coefficient. For the tuning of resonance wavelength, only the active region of microring is heated electrically to induce temperature-dependent changes in the refractive indices.
%%%%%%%%%%%%%%%%%%%%%%%%%%%%%%%%%%%%%%%% Design And Simulations %%%%%%%%%%%%%%%%%%%%%%%%%%%%%%%%%%%%%%%%
%%%%%%%%%%%%%%%%%%%%%%%%%%%%%%%%%%%%%%%%%%%%%%%%%%%%%%%%%%%%%%%%%%%%%%%%%%%%%%%%%%%%%%%%%%%%%%%%%%%%%%%%
\section{Design and Numerical Analysis}
For designing the tunable optical filter the considered cross-section of ring and bus waveguides are shown in Fig.~\ref{fig1}(b). Both waveguides are placed on top of a 2 $\mu$m thick $SiO_{2}$ layer followed by an 80 nm thick Si layer. The ring has an inner radius of 5 $\mu$m, and separation between the bus and ring waveguides is denoted as g. As illustrated in Fig.~\ref{fig1}(c), in the active region of microring, a 2 $\mu$m long and 20 nm thick GST layer is embedded in a partially etched portion of the Si microring of height 200 nm. The Si layer beneath the GST is n-type doped to provide conductive passage for the input electrical signal, applied for the Joule heating of the GST layer. Potentially CMOS compatible indium-tin-oxide (ITO) with thickness of 50 nm is used as an electrode material \cite{b17}. The critical coupling between the microring and bus waveguide is achieved at the coupling gap of 100 nm and the same is shown in Fig.~\ref{fig1}(d). The device is modelled using co-simulation module of CST microwave studio software. The various equations governing three types of simulations (electrical, thermal and optical) in the co-simulation module are described in following paragraphs.\\
Electrical equations are solved to obtain the current and voltage distribution in the device. The applied voltage pulses generate current which flow through the hybrid GST-Si part and causes Joule heating, resulting in the temperature rise of GST and its surrounding region. The equations solved in the electrical model are following:
\begin{table}
\centering
\caption{Materials coefficients used for the thermal simulations}
%\begin{indented}
%\lineup
\label{tab:1}
\begin{tabular}{ c c c c c }

\hline
Coefficients & $SiO_{2}$ & Si & ITO & a-GST \cr
\hline
Refractive index & 1.45 & 3.48 &  1.6+i0.35 & 4.62+i0.12 \cr
Density   & 2202 & 2329 & 6800 & 5900 \cr
(Kg/$m^{3}$) & &  & &  \cr
Specific heat  & 746 & 713 & 1340 & 212 \cr
 (J/Kg K) & &  & &  \cr
 Electrical cond. & $10^{-14}$& $10^{3}$ &  8.3$\times$ $10^{3}$ & 3\\
(S/m)   & &  & &   \cr
Thermal cond. & 1.38 & 140 & 5 & 0.19 \cr
(W/mK) & &  & &   \cr
\hline
\end{tabular}
\end{table}
%\end{indented}

\begin{figure}
	\centering
		\includegraphics[scale=.5]{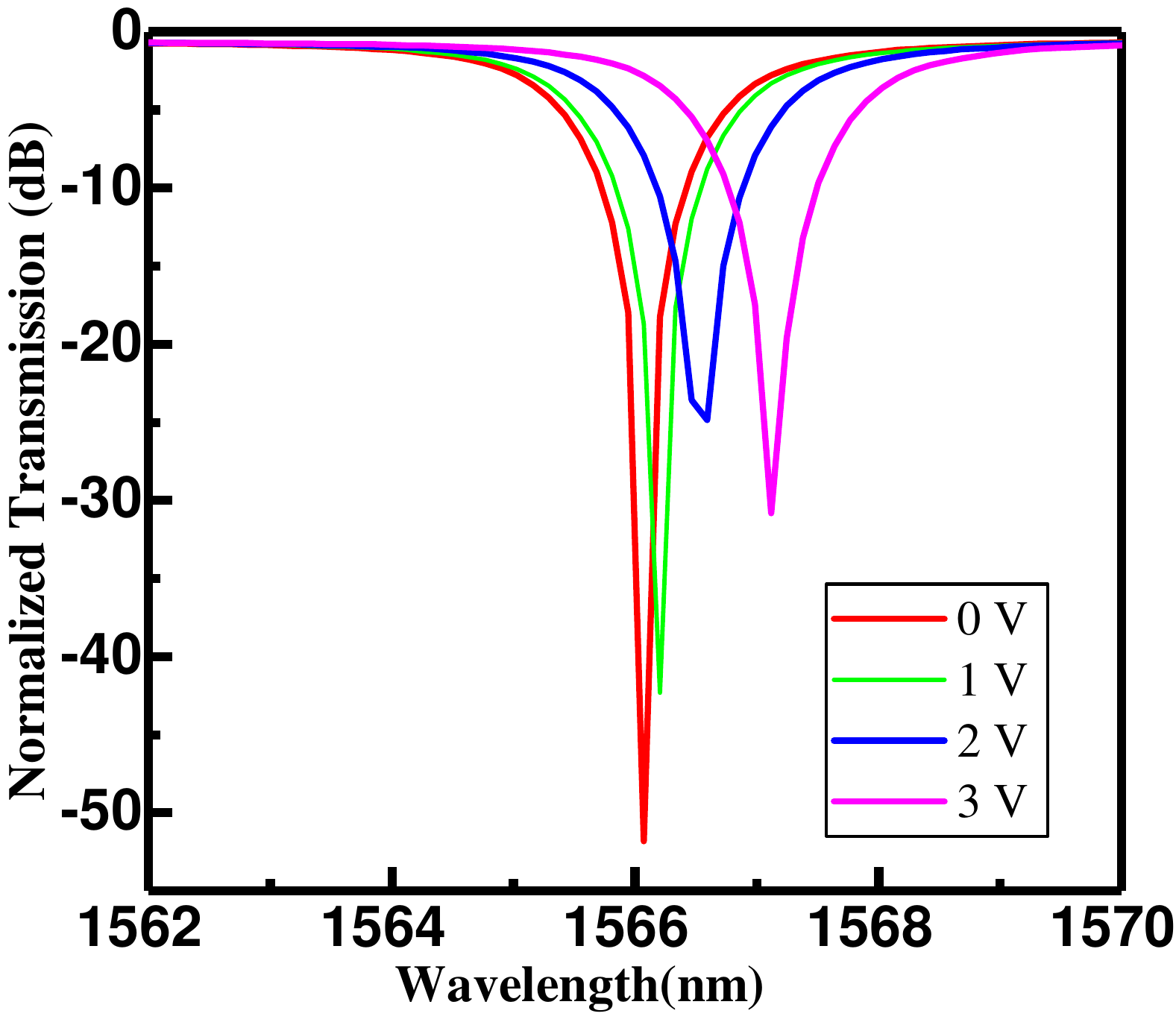}
	\caption{Normalized output transmission of the proposed tunable optical filter for different applied voltages.}
	\label{fig2}
\end{figure}
\begin{eqnarray}
\vec \nabla \cdot [\sigma (x,y,z,t) \vec \nabla V] = 0\\
\vec j = \sigma(x,y,z,t) \vec E
\end{eqnarray}
In eq (1) and (2), $\sigma$, $V$, $j$ and E denote electrical conductivity, voltage, current density and electric field respectively. Electrical conductivity of GST in amorphous phase is set to be 3 S/m \cite{b18}.\\
In thermal model, the temperature distribution in the active region is obtained by solving the transient thermal equations given below:
\begin{eqnarray}
\vec \nabla \cdot [k(x,y,z) \vec \nabla T] + Q = C_v (\frac{d T}{d t})\\
Q = \sigma E^{2}
\end{eqnarray}
 In eq (3) and (4), k, T, $C_{v}$, Q, E, and $\sigma$ denote thermal conductivity, temperature, volumetric heat capacity, Joule heat generation, electric field, and electrical conductivity respectively. The material parameters used in thermal simulations are presented in Table \ref{tab:1}.\\
 In the optical model, Maxwell's equations are solved to obtain 3D electrical field distribution and transmission through the device and same are as follows: 
\begin{eqnarray}
\vec\nabla \times \vec H = \vec j + \epsilon(z) \frac{d \vec E}{dt}\\
\vec\nabla \times \vec E = - \mu_{0} \frac{d \vec H}{dt}\\
%\epsilon = \epsilon' + \iota \epsilon''\\
n = n_r + \iota k 
\end{eqnarray}
The temperature dependent variation in the amorphous GST refractive index is included in the model via the thermo-optic coefficient given by \cite{b16}:
\begin{eqnarray}
\left(\frac{dn}{dT}\right)_{a-GST} = 1.1 \times 10^{-3} K^{-1}\\
\left(\frac{dk}{dT}\right)_{a-GST} = 4.1 \times 10^{-4} K^{-1}
\end{eqnarray}
For Silicon, the thermo-optic coefficient is taken as \cite{b15}:
\begin{equation}
\left(\frac{dn}{dT}\right)_{Si} = 1.8 \times 10^{-4} K^{-1}    
\end{equation}

%In the case of GST, the thermo-optical effect, the output of the device changes with increasing temperature, change in the real part of refractive index leads to a phase shift, while the change in the imaginary part of the refractive index leads to attenuation. 
The various material parameters used for the simulations are listed in Table \ref{tab:1} \cite{b19,b20,b21}.
\section{Results and Analysis}
\begin{figure}
	\centering
		\includegraphics[scale=.5]{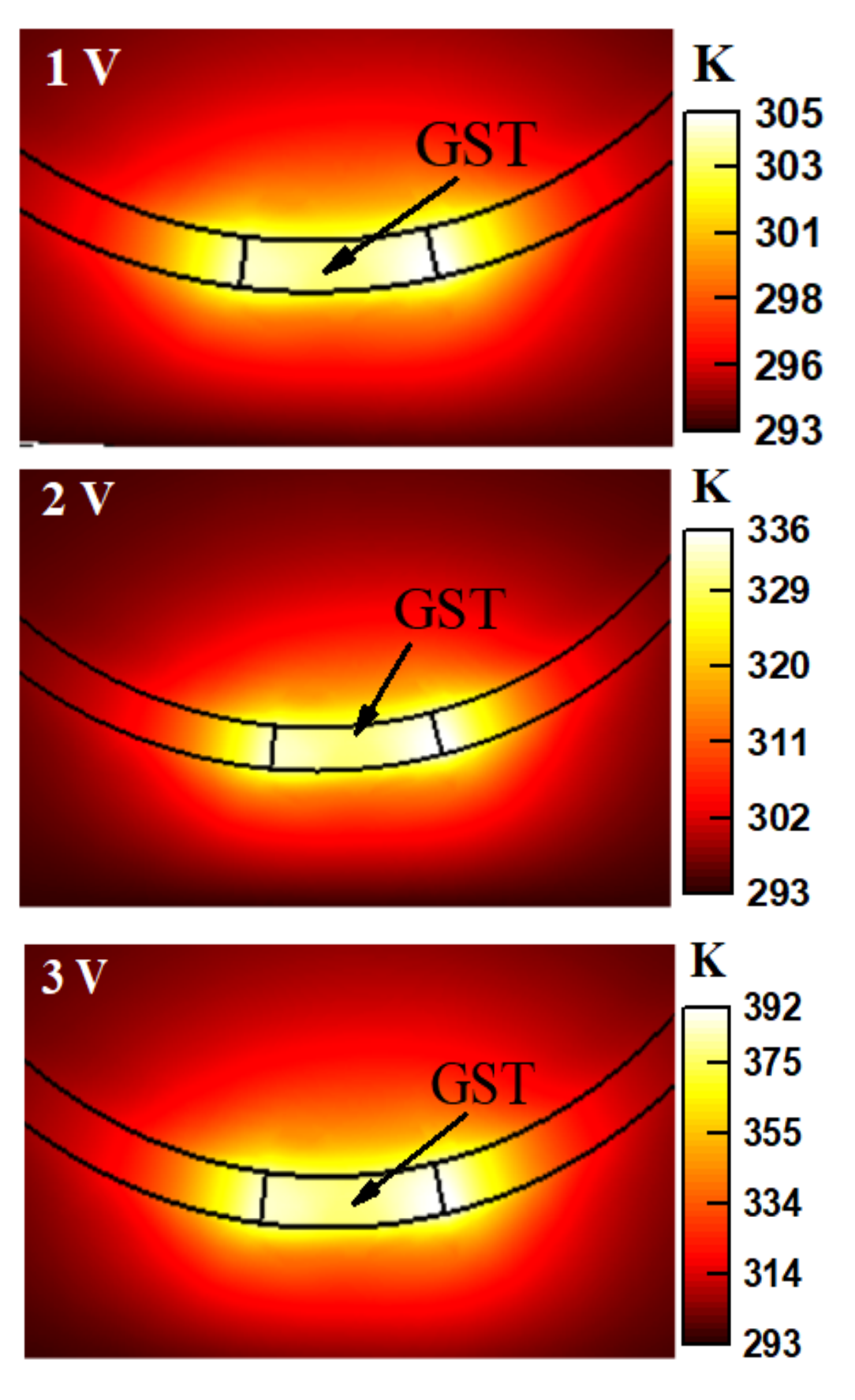}
	\caption{The temperature raise in the active part of the microring resonator for different applied voltages.}
	\label{fig3}
\end{figure}

\begin{figure}
	\centering
		\includegraphics[scale=.5]{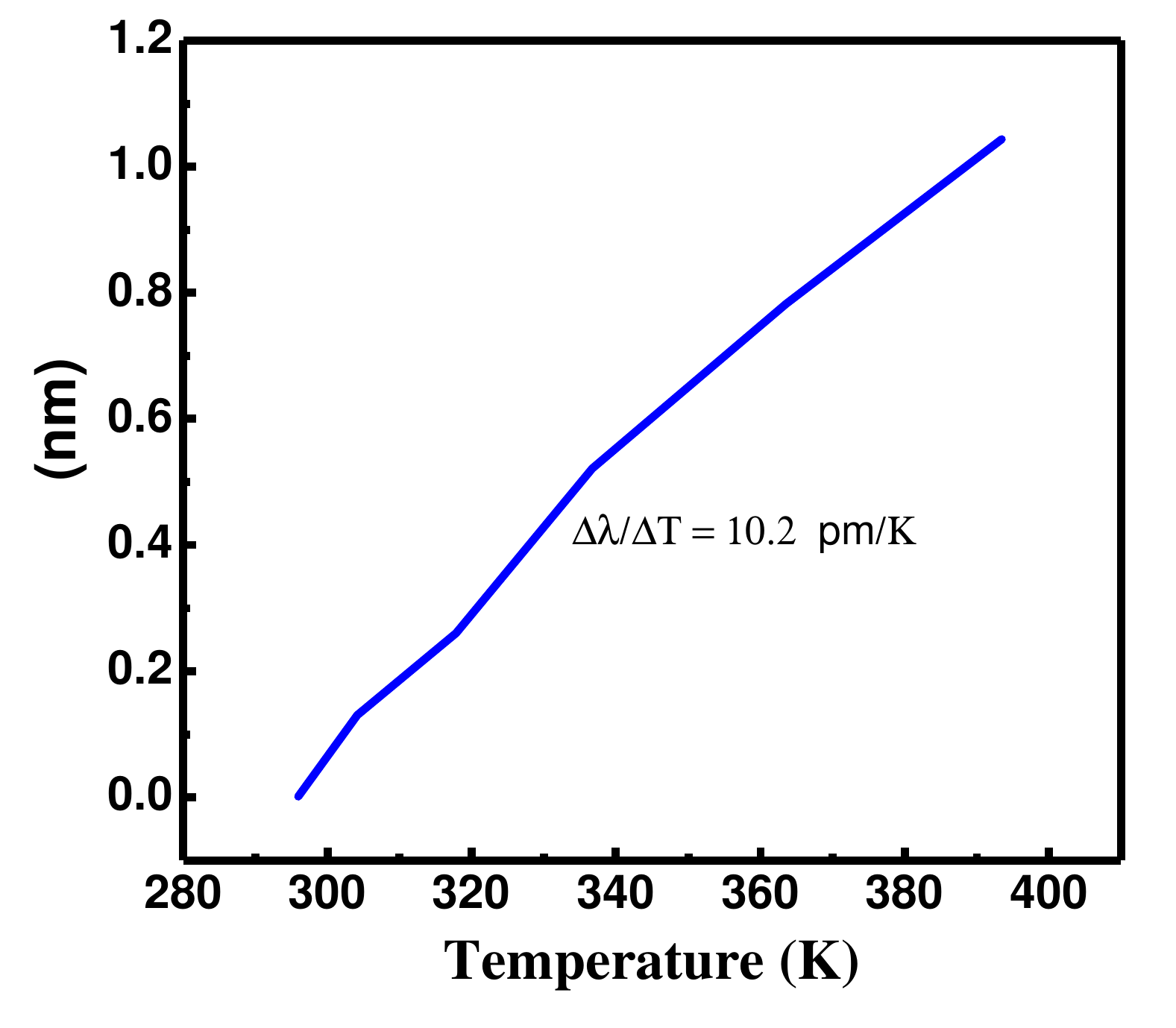}
	\caption{The shift in resonance wavelength with the increase in the temperature of the active part of the microring.}
	\label{fig4}
\end{figure}

The device is designed to be critically coupled when there is no voltage applied. At critical coupling,  output transmission has an extinction ratio of $\sim$51 dB as shown by the red line in Fig.~\ref{fig2}. 
%Such a high extinction ratio at resonance wavelength is due to the critical coupling between the bus and ring waveguides with a coupling gap of 100 nm. 
As microring is heated above room temperature, the resonant wavelength is red-shifted with decrease in resonance dip as illustrated by the transmission spectrum shown in Fig.~\ref{fig2}. The temperature raise of the GST and it's surrounding increases both the real and imaginary part of the refractive index. The change in real part of the refractive index results in a red shift of resonance wavelength while increase in absorption results in the change of extinction ratio of resonance dip. Additional thermo-optic effect in silicon is also expected to affect the output transmission. However, this effect is substantially weaker than the much larger optical response of the GST. The red shift in resonance wavelength increases linearly with the applied voltage while the decrease in resonance dip depends upon temperature variations of the active region. The resonance depth becomes shallow due to increase in absorption loss of GST with the increase in applied voltage.\\
To obtain the temperature distribution of device as a function of applied voltage, we performed the electro-thermal simulations of the device. The voltage applied through the ITO electrodes generates current through the hybrid GST-Si part and causes Joule heating. The thermal profiles of hybrid part of the ring with the cross section of the GST layer are shown in Fig.~\ref{fig3}. A voltage pulse of 1 V applied for 100 ns time duration causes maximum temperature rise of 308 K in the hybrid part of microring waveguide. This leads to decrease in the resonance dip to 42.3 dB, and a red shift of 130.4 pm in resonant wavelength. The shift in the resonant wavelength as a function of temperature of active region is shown in Fig. \ref{fig4}. Our simulations reveal a sizeable shift $\Delta \lambda$ = 1.04 nm in the resonance wavelength, with an integrated 2 $\mu$m long and 20 nm thick GST layer embedded in a portion of the microring waveguide. The tuning of resonance wavelength is limited due to the requirement of maintaining the GST temperature below glass transition temperature (413 K) \cite{b22}. Above this temperature GST may transform into the crystalline phase. To the best of authors knowledge this is the first work reported for electrically tunable Si photonic filter enabled by embedding GST in microring resonator.
%Figure 8a shows the transmission spectra for an ultra-compact Si-VO2 hybrid micro-ring resonator, R = 1.5 μm, with an integrated ~500 nm long and ~70 nm thick VO2 patch coating a portion of the ring waveguide.
%These measurements reveal a sizeable shift Δλ = –1.26nm in the resonance wavelength, coinciding with an optical modulation greater than 10dB at the initial resonance position, λ = 1568.78nm.
% However, additional effects are also expected to be present during this experiment, including dependence on: (1) the thermo-optic (TO) effect in silicon, Δn/ΔT = +1.86×10-4/K [3], and (2) the free-carrier index (FCI) [4]. These two effects are substantially weaker than the much larger optical response of the VO2 and, in this experiment, also carry opposite signs.

%%%%%%%%%%%%%%%%%%%%%%%%%%%%%%%%%%%%%%%%%%%%%%%%%%%%%%%%%%%%%%%%%%%%%%%%%%%%%%%%%
%%You are recommended to conform your choice to the journal you are submitting to.
%%%%%%%%%%%%%%%%%%%%%%%%%%%%%%%%%%%%%%%%%%%%%%%%%%%%%%%%%%%%%%%%%%%%%%%%%%%%%%%%%
\section{Conclusions}
 We proposed and investigated an ultra-compact and reconfigurable optical filter with high extinction ratio using microring resonator based on  GST-Si hybrid platform. By using GST with microring resonator, and exploiting its high thermo-optic coefficient in amorphous phase, we successfully achieved the tuning of the resonance wavelength. A resonance wavelength shift of more than 1000 pm with extinction ratios in the range of 20-41 dB are obtained. The voltage required to tune the resonance wavelength by more than 1000 pm is 3 V only. %Our simulation results reveal a sizeable shift of 1.04 nm in resonance wavelength with small active GST volume of  2 $\mu$m $\times$ 0.480 $\mu$m $\times$ 0.02 $\mu$m (length $\times$ width $\times$ height). 
This optical filter can be utilized in large scale complex reconfigurable systems such as reconfigurable optical add/drop multiplexer (ROADM). Also, optical switching can be performed in the proposed device by using the add-drop configuration of microring resonator. In such configuration, one may switch from the on resonance (amorphous GST) to the off-resonance (crystalline GST) and vice versa to achieve switching of the optical signal.
%%%%%%%%%%%%%%%%%%%%%%%%% Conculsions %%%%%%%%%%%%%%%%%%%%%%%%
%%%%%%%%%%%%%%%%%%%%%%%%%%%%%%%%%%%%%%%%%%%%%%%%%%%%%
%%%%%%%%%%%%%%%%%%%%%%%%%%%%%%%%%%%%%%%%%%%%%%%%%%%%%%%%%%%%%%%%%%%%%%%%%%%%%%%%%
%%You are recommended to conform your choice to the journal you are submitting to.
%%%%%%%%%%%%%%%%%%%%%%%%%%%%%%%%%%%%%%%%%%%%%%%%%%%%%%%%%%%%%%%%%%%%%%%%%%%%%%%%%
\section{Acknowledgements}
This work is supported by the research projects funded by the IIT Roorkee under FIG scheme; and Science and Engineering Research Board (SERB), Department of Science and Technology, Govt. of India under Early Career Research award scheme (project File no. ECR/2016/001350)
\bibliographystyle{unsrt}

\end{document}